\documentclass[aip,APL, amsmath,amssymb,reprint]{revtex4-1}
\usepackage{graphicx}
\usepackage{dcolumn}
\usepackage{bm}
\usepackage{subfigure}
\usepackage{epstopdf}
\DeclareMathAlphabet{\mathpzc}{OT1}{pzc}{m}{it}

\begin{document}

\title[Graphene with adatoms: tuning the magnetic moment]{Graphene with adatoms: tuning the magnetic moment with an applied voltage}

\author{N. A. Pike}
\email{Pike.55@osu.edu}
\author{D. Stroud}%
 \email{Stroud@physics.osu.edu}
\affiliation{ Department of Physics, The Ohio State University, Columbus, OH 43210}%

\date{\today}

\begin{abstract}

We show that, in graphene with a small concentration of adatoms, the total magnetic moment $\mu_T$ can be switched on and off by varying the Fermi energy $E_F$, either by applying a gate voltage or by suitable chemical doping.  Our calculation is carried out using a simple tight-binding model described previously, combined with a mean-field treatment of the electron-electron interaction on the adatom.    The values of $E_F$ at which the moment is turned on or off are controlled by the strength of the hopping between the graphene sheet and the adatom, the on-site energy of the adatom, and the strength of the electron-electron correlation energy  U.   Our result is in qualitatively consistent with recent experiments by Nair {\it et al.} [Nat.\ Commun.\ {\bf 4}, 2010 (2013)]. 
\end{abstract}

\pacs{73.20.At, 73.22.Pr, 75.70.Ak}

\maketitle
The two-dimensional structure of graphene and the Dirac-like dispersion relation of its electrons are the origin of many unusual properties,  which may lead to novel electronic or spintronic applications~\cite{pesin,tombros,swartz}.  For example, adatoms on graphene may develop magnetic moments which can be manipulated by an applied electric field in a manner similar to its other electric and optical properties~\cite{garnica,geim,hu,yun}.  Recent experimental work by Nair {\it et al.} shows that both $sp^3$ defects and vacancies in graphene possess  magnetic moments that can be switched on and off by chemical doping~\cite{nair}.  

In a recent paper, we showed, using a tight-binding model, that certain non-magnetic adatoms, such as H, can create a non-zero magnetic moment on graphene~\cite{pike}.  In this Letter, we extend our calculation to show that the magnetic moment of graphene with adatoms can be switched on and off by varying the Fermi energy.  This Fermi energy can be controlled, in practice, by an applied voltage; it can also be tuned by suitable chemical doping of the graphene, as in the experiments of Ref.~\citenum{nair}.  Our calculations show that the onset and turn-off of the magnetic moment depend on the parameters characterizing the adatom, such as the hopping strength between the adatom and  graphene, the on-site energy, and the electron-electron correlation energy. 

In the following section we briefly review our model, and its solution via mean field theory.  The model is appropriate when the adatom lies atop one of the $C$ atoms in graphene (the so-called T site), as is the case for adsorbed H and several other adatom species\cite{nakada,chan}.  In section $2$, we present  numerical results showing the dependence of the magnetic moment on Fermi energy for various model parameters.   In section $3$ we give a concluding discussion.

\section{Tight-Binding Model}
We consider a tight-binding Hamiltonian to model the graphene-adatom system.  The graphene part of the Hamiltonian, denoted $H_0$, is written in terms of the creation and annihilation operators for electrons of spin $\sigma$ on a site in the $n^{th}$ primitive cell~\cite{pike}.  Denoting the creation (annihilation) operators for the $\alpha$ and $\beta$ sub-lattices by  $a_{n\sigma}^\dag$  ($a_{n\sigma}$) and $b_{n\sigma}^\dag$ ($b_{n\sigma}$), we write $H_0$ for nearest-neighbor hopping on graphene as $H_0 = \sum_{{\bf k},\sigma}H_{0,{\bf k},\sigma}$, where
\begin{equation}\label{eq:CCham}
H_{0,{\bf k},\sigma} =  -t({\bf k}) a^\dagger_{{\bf k},\sigma} b_{{\bf k,\sigma}}-t^*({\bf k})a_{{\bf k},\sigma}b^\dagger_{{\bf k},\sigma}
\end{equation}
and $\sigma = \pm 1/2$. We note that here and in subsequent equations the notation of Ref.~\citenum{pike} is used. 

 In Eq. (\ref{eq:CCham}), $t({\bf k})=t\left[ 1+ 2\exp\left(\frac{3 i k_x a_0}{2}\right) \cos\left(\frac{\sqrt{3} k_y a_0}{2}\right)\right]$, where $a_0 =1.42 \AA$ is the nearest-neighbor bond length for graphene, and $t$  is the hopping energy between nearest neighbor carbon atoms (for graphene $t = 2.8 \ eV$)~\cite{Rakhmanov12,Yazyev2007}.  The extra part of the Hamiltonian due to an adatom at a $T$ site may be written in real space as
\begin{equation}\label{eq:hireal}
H_I = \epsilon_0 \sum_\sigma h_{0,\sigma}^\dag h_{0,\sigma} - t^\prime\sum_\sigma \left(h_{0,\sigma}^\dag a_{0,\sigma} + h_{0,\sigma}a_{0,\sigma}^\dag\right),
\end{equation}
where $h_{0,\sigma}^\dag$ and $h_{0,\sigma}$ are creation and annihilation operators for an electron of spin $\sigma$ at the site of the adatom, $\epsilon_0$ is the on-site energy of an electron on that site (relative to the Dirac point of the pure graphene band structure), and $t^\prime>0$ is the energy for an electron to hop between the adatom and the carbon atom at the site $0$ of the $\alpha$ sub-lattice.   
In terms of Bloch eigenstates of $H_0$,  
\begin{eqnarray}\label{himp}
H_I &=& \epsilon_0 \sum_\sigma h_{0,\sigma}^\dag h_{0,\sigma}  \\ \notag & -& \frac{t^\prime}{\sqrt{2N}} \sum_\sigma\left[ h_{0,\sigma}^\dag \sum_{\bf k}e^{-i\phi_{\bf k}}(\gamma_{{\bf k},\sigma,1} +\gamma_{{\bf k},\sigma,2}) + h.c.\right].
\end{eqnarray} 
\begin{widetext}
Here $\gamma_{{\bf k},\sigma,i}$ (i=1,2) is the annihilation operator for a Bloch electron of wave vector ${\bf k}$, spin $\sigma$, and within the $i^{th}$ band.  $\phi_{\bf k}$ is a phase angle given in Ref.~\citenum{pike} and for a hydrogen adatom  we take $\epsilon_0 = 0.4 \ eV$ and $t' = 5.8 \ eV$~\cite{Rakhmanov12}.  

To calculate the magnetic moment we add a Hubbard term $H_U$, which acts only on the adatom. Within the mean-field approximation $H_U$ is given by
\begin{equation}\label{MFT_U}
H_U  \sim U \left[h_{0\uparrow}^\dagger h_{0\uparrow} \langle n_{0\downarrow}\rangle+h_{0\downarrow}^\dagger h_{0\downarrow} \langle n_{0\uparrow}\rangle- \langle n_{0\uparrow}\rangle \langle n_{0\downarrow}\rangle\right].
\end{equation}
where  $\langle n_{0,\sigma}\rangle$ is the average number of electrons with spin $\sigma$ on a $T$ site of the $\alpha$ sub-lattice.   For a hydrogen adatom we take $U$ to be the difference between the ionization potential and the electron affinity, which gives $U \sim 12.85 eV = 4.59t$~\cite{pariser,lykk}.

The total mean-field Hamiltonian consists of the sum of Eqs.\ (\ref{eq:CCham}), (\ref{himp}), and (\ref{MFT_U}), which is quadratic in the electron creation and annihilation operators and therefore readily diagonalized.   The total spin-dependent densities of states $\rho_{tot,\sigma}(E)$  can then be calculated,  given the values of $\langle n_{0,\sigma}\rangle$.   The result is~\cite{pike}
\begin{equation}\label{eq:total_dos}
\rho_{tot,\sigma}(E) = N\rho_{0}(E) -\frac{1}{\pi}\left[\mathrm{Im}\left( \frac{d}{dz}\mathrm{ln}[z-\epsilon_0-U\langle n_{0,-\sigma}\rangle-\frac{t^{\prime 2}}{2N}{\cal G}_0(z)]\right)\right]_{z = E + i0^+},
\end{equation}
where $\rho_0(E)$ is the density of states per spin and per primitive cell of pure graphene, $N$ is the number of primitive cells, and ${\cal G}_0(z)$ the corresponding Green's function, as defined in Ref.~\citenum{pike}.  The quantities $\langle n_{0,\sigma}\rangle$ are obtained by integrating the local spin-dependent density of states on the adatom, $\rho_{00,\sigma}(E)$, from the bottom of the valence band up to the Fermi energy.  $\rho_{00,\sigma}(E)$ is, in turn, given by 
\begin{equation}\label{eq:local_dos}
\rho_{00,\sigma}(E) = -\frac{1}{\pi}\mathrm{Im}\left(\frac{1}{z-\epsilon_0-U\langle n_{0,-\sigma}\rangle - \frac{t^{\prime,2}}{2N}{\cal G}_0(z)}\right)_{z = E + i0^+}.
\end{equation}
\end{widetext}

The total magnetic moment $\mu_T$ can be calculated as a function of the Fermi energy $E_F$ from the expression
\begin{equation}
\label{eq:magmom}
\mu_T(E_F)  = \mu_B \int_{-3t}^{E_F}\left[\rho_{tot,\uparrow}(E)-\rho_{tot,\downarrow}(E)\right]dE,
\end{equation}
where $\mu_B$ is the Bohr magneton. 

Using these results, we can numerically calculate $\mu_T(E_F)$  for various choices of tight binding parameters.  This is done iteratively, for a given choice of $E_F$,  as follows.  First, we make initial guesses for $\langle n_{0,\sigma}\rangle$.  Next, we integrate the spin-dependent  local density of states up to the Fermi energy using Eq.\ (\ref{eq:local_dos}) and the initial guesses for $\langle n_{0,\sigma}\rangle$.    This gives the next iteration of $\langle n_{0,\sigma}\rangle$.   We iterate until the changes in $\langle n_{0,\sigma}\rangle$ in two successive cycles are less than $0.001$.   Finally, using the converged values we calculate the total magnetic moment from Eq.\ (\ref{eq:magmom}).

\begin{figure}[t]\centering
\begin{minipage}[b][0.6\textheight][s]{0.5\textwidth}
        \includegraphics[height=0.2\textheight,width=0.8\textwidth]{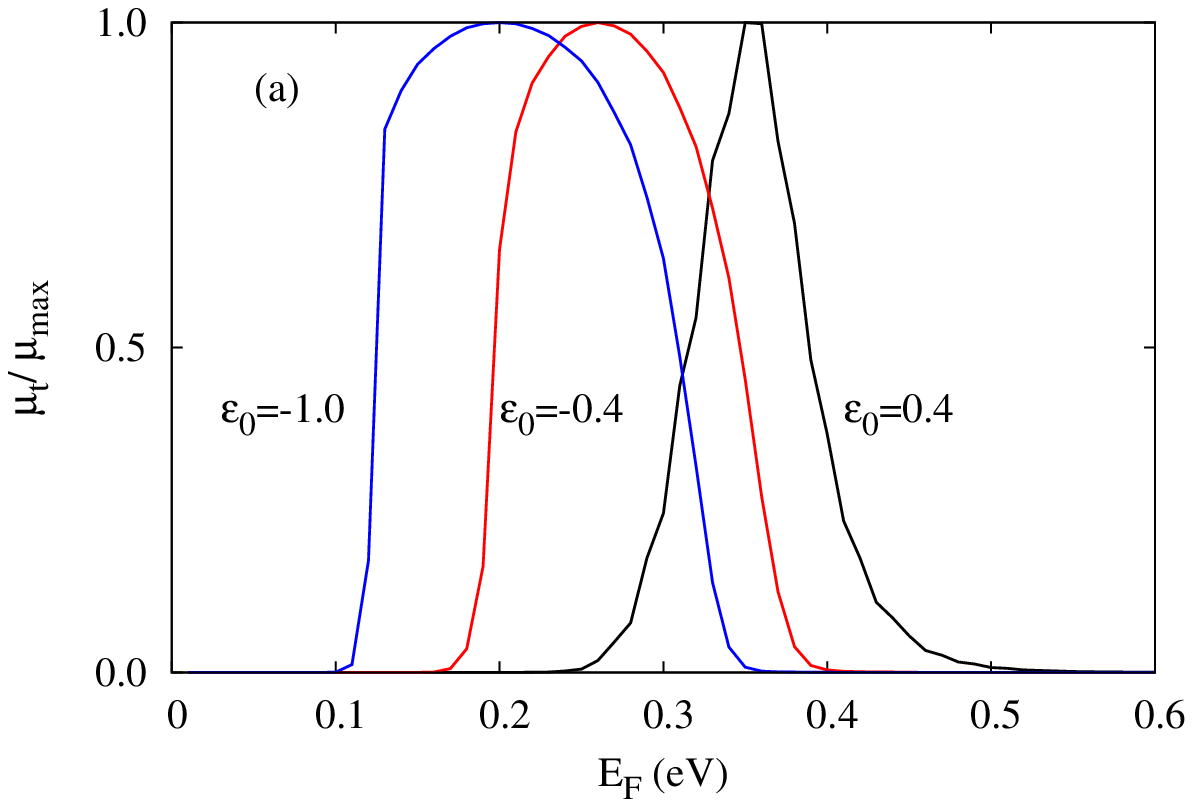}
\vfill
         \includegraphics[height=0.2\textheight,width=0.8\textwidth]{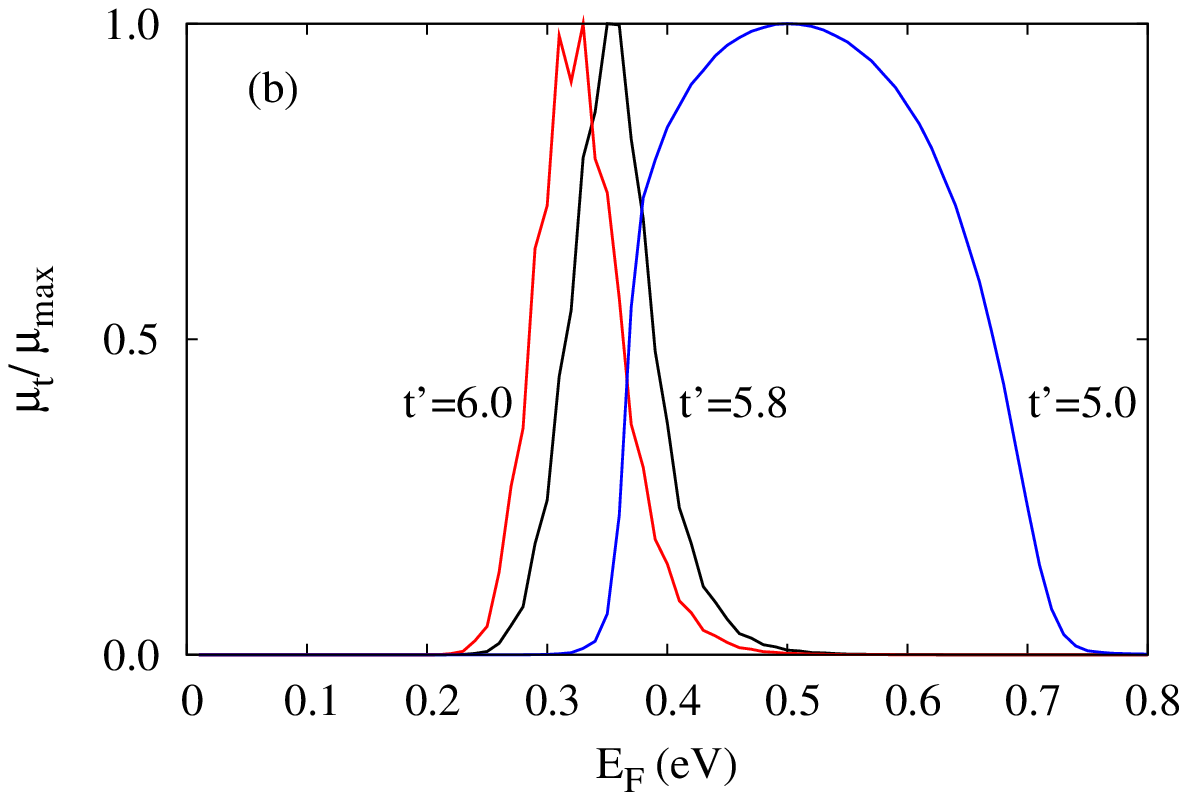}
\vfill
        \includegraphics[height=0.2\textheight,width=0.8\textwidth]{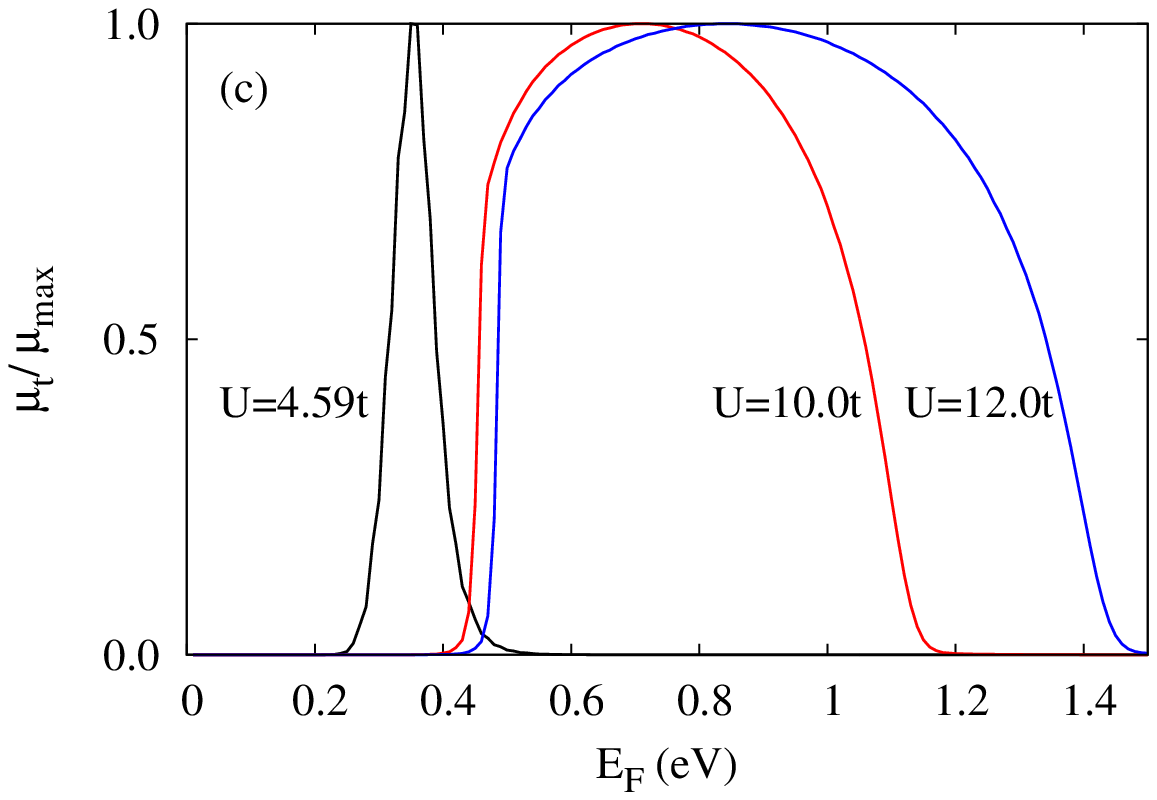}
\end{minipage}
\caption{(Color online) Total magnetic moment verse Fermi energy for an adaom on graphene calculated using Eq. (\ref{eq:magmom}).  In each case, all the parameters but one
are appropriate to H on graphene, as given in the text.  In (a) each curve represents a different on-site energy $\epsilon_0$ (given in eV in the legend).   In (b), each curve represents to a different hopping energy $t'$ (in eV), and in (c) each curve represents a different Hubbard energy $U$ (in eV). The maximum calculated magnetic moment in each Figure is noted within the text. }
\label{fig1}
\end{figure}

\section{Numerical Results}

We have carried out these calculations as a function of $E_F$ for a variety of values of $\epsilon_0$, $t^\prime$, and $U$.   In every case, we find that the total magnetic moment is non-zero only when $E_F$ lies within a limited range.  We write this condition as $E_\ell< E_F < E_\mathpzc{u}$ where $E_\ell$ and $E_\mathpzc{u}$ are the lower and upper energies within which $\mu_T \neq 0$.   Within this range, the magnitude of $\mu_T$ is controlled by varying the Fermi energy $E_F$, typically by applying a gate voltage to the sample which interacts with the conduction electrons via the electric field effect~\cite{geim}.  If the graphene-adatom system is neutral and no voltage is applied, $E_F$ will be constrained to have a particular value controlled by the charge neutrality condition.  Introducing a gate voltage will shift $E_F$ from this neutral value (denoted $E_{F_0}$) and hence change the magnetic moment.

As an illustration of this picture, we show in Fig.~\ref{fig1} the total magnetic moment $\mu_T(E_F)$ under various conditions.  In each case, we assume all parameters but one are those thought to describe H on graphene, and vary the remaining parameter~\cite{Rakhmanov12,Yazyev2007,pariser,lykk}.  We assume a single H adatom is placed on a graphene sheet containing $N=500$ carbon primitive cells,  giving 1 H atom per 1000 C atoms.   In Fig.~\ref{fig1}(a), we assume that $t$ and $t^\prime$ are those of the H-graphene system, while each curve represents a different value of the on-site energy $\epsilon_0$.   We find that, when $E_{F_0} > \epsilon_0$  and we allow $\epsilon_0$ to become much less then $ E_{F_0}$ that the onset energy $E_\ell \rightarrow 0$ , whereas the upper cutoff energy $E_\mathpzc{u}$ shows only a minimal dependence on $E_F$. 

 In  Fig.~\ref{fig1}(b) we plot $\mu_T(E_F)$  versus $E_F$ for several values of $t^\prime$, with other parameters the same as the H-graphene system.   As $t^\prime$  is reduced, the upper energy cutoff, $E_\mathpzc{u}$, also decreases, whereas the lower energy onset remains approximately unchanged at $E_\ell=0.3 eV$, independent of $t^\prime$.  In Fig.~\ref{fig1}(c),   we plot $\mu_T(E_F)$  versus $E_F$ for several values of the electron-electron energy $U$, assuming the other parameters same as the H-graphene system.  As $U$ increases, so do both $E_\ell$ and $E_\mathpzc{u}$.  As $U \rightarrow \infty$, $E_\mathpzc{u} \rightarrow \infty$ as well, i.\ e., $\mu_T$ persists no matter how large $E_F$ in this case.

For all parameters we have considered, we find that the magnitude of the magnetic moment $\mu_T < \mu_B$.    In Fig.~\ref{fig1}(a). the maximum value of $\mu_T \approx 0.1 \mu_B$, corresponding to $\epsilon_0 =-0.4\ eV$ and $-1.0\  eV$. 
 In Fig.~\ref{fig1}(b) the maximum value of $\mu_T \approx 0.14 \mu_B$ for $t^\prime = 5.0 eV$, while in Fig.~\ref{fig1}c a maximum value of $\mu_T \approx 0.2 \mu_B$ for  $U= 12.0 t$. In previous work~\cite{pike}, it has been shown that $U\rightarrow \infty$, $\mu_T \rightarrow \mu_B$.

\section{Discussion and Conclusions}

The numerical results of Fig.~\ref{fig1} can be qualitatively understood as follows.  If the system of graphene plus adatom has a net magnetic moment, then the partial densities of states $\rho_\uparrow(E)$ and $\rho_\downarrow(E)$ will be different.  The total magnetic moment is then obtained by integrating these two densities of states up to $E_F$.   If $E_F$ lies below the bottom edge of the  lower band, there will be no net magnetic moment.  The moment becomes non-zero at the energy $E_F = E_\ell$ when $E_F$ moves above the bottom of the lower band.   It reaches its maximum when $E_F$  lies somewhere between the peaks in $\rho_\uparrow(E)$  and $\rho_\downarrow(E)$ (assuming the two bands overlap),  and thereafter decreases until it becomes zero at $E_F=E_\mathpzc{u}$, the energy above which both sub-bands are filled.  Thus, for any choice of the parameters $\epsilon_0$, $t^\prime$, and $U$, there should be a finite range of $E_F$, within which $\mu_T \neq 0$. 

Of course, this description is an oversimplification because the self-consistently determined $\rho_\uparrow(E)$ and $\rho_\downarrow(E)$  depend on the quantities $\langle n_{0,\sigma}\rangle$ ($\sigma = \pm 1/2$), which themselves depend on $\mu_T$.  However, even with the oversimplification, the qualitative description remains correct.   The maximum value of $\mu_T$ depends, in part,  on how much the two sub-bands overlap.  If the overlap is large, maximum of $\mu_T$ will be small, everything else being equal, while a small overlap will tend to produce a larger $\mu_T$.     Another reason why $\mu_T$ is reduced below $\mu_B$ is that there is generally a large electron transfer from the adatom onto the graphene sheet\cite{pike}.  We also note that the energy range, $E_\mathpzc{u} - E_\ell$, where $\mu_T \neq 0$  is approximately equal to the width of the extra density of states due to the adatom.

The Fermi energy $E_F$ can be controlled experimentally in several ways.   One is to apply a suitable gate voltage $V$, which raises or lowers $E_F$ by an amount $eV$, where $e$ is the magnitude of the electronic charge. Another is by chemical doping: if one adds or subtracts charge carriers to the graphene-adatom system by doping with suitable molecules, this will also raise or lower $E_F$ as was done in Ref.~\citenum{nair}. One  could also add a small number of vacancies in the graphene, as also done in Ref.~\citenum{nair}.  This will reduce the number of charge carriers and hence lower $E_F$.  Of course, vacancies would also change the graphene density of states; so the present calculations would have to be modified to treat this situation.   

The model used here treats only the effects of adatoms on graphene, but it does qualitatively reproduce the upper energy cutoff found in the experiments of Ref.~\citenum{nair}.  For example, Nair {\it et al.}~\cite{nair} found an upper cutoff of around $E_\mathpzc{u} = 0.5 eV$, which we can approximately obtain by assuming an on-site energy $\epsilon_0 = 0.4 eV$, $t' = 5.8 eV$ and $U = 4.59 t$.   However, our model does not account for the onset energy found in Ref.\ \cite{nair} of $E_F\sim 0$ since in our model the bottom edge of the lower band occurs at $E_\ell >0 eV$. 

In summary, by using a simple tight-binding model of adatoms on graphene we are able to calculate the total magnetic moment of graphene with a small concentration adatoms as a function of $E_F$.   The model is expected to apply to the case of H adatoms, but could also be applicable to other adatom species, characterized by different model parameters.  The Fermi energy $E_F$ can be controlled experimentally by a suitable gate voltage.   Our results show that, for realistic tight-binding parameters ($\epsilon_0 = 0.4 eV, t' = 5.8 eV, U = 4.59t$), the magnetic moment can be switched off at a relatively low voltage ($eV \sim 0.5eV$),  in rough agreement with the experiments of Ref.~\citenum{nair}.    These results are potentially of much interest since they suggest that the magnetic moment of graphene with adatoms can be electrically controlled.

\section{Acknowledgments}
This work was supported by the Center for Emerging Materials at The Ohio State University, an NSF MRSEC (Grant No.\ DMR0820414).   We thank R.\ K.\ Kawakami for helpful discussions.


\begin{thebibliography}{99}
\bibitem{pesin} D. Pesin and A. H. MacDonald,
Nat. Maters. {\bf 11} 409-416 (2012).
\bibitem{tombros} N. Tombros, C. Jozsa, M. Popinciuc, H.\ T.\ Jonkman, and B.\ J.\ van Wees,
Nature {\bf 448}, 06037 (2007).
\bibitem{swartz} K.\ M.\  McCreary,  A.\ G.\  Swartz, W.\  Han, J.\  Fabian, and  R.\ K.\ Kawakami, 
Phys.\ Rev.\  Lett.\ {\bf 109}, 186604 (2012).
\bibitem{garnica}M.  Garnica, D. Stradi, S. Barja, F.  Calleja, C.  Diaz, M. Alcami, N. Martin, A. L. Vazquez de Parga, F.  Martin, R.  Miranda. 
Nat. Phys. {\bf 9} 368-374 (2013).
\bibitem{geim}A. K. Geim. 
Science {\bf 324} 1530-1534 (2009).
\bibitem{hu}F. M. Hu, T.  Ma, H. Lin,and J. E. Gubernatis.
Phys. Rev. B. {\bf 84}, 075414 (2011).
\bibitem{yun}K-H. Yun, M. Lee, and Y-C. Chung.
J. Magn. Magn. Mater. {\bf 392} 93-96 (2014).
\bibitem{nair} R. R Nair, I-L Tsai, M. Sepioni, O. Lehtinen, J. Keinonen, A.V Krasheninnikov, A. H Castro Neto, M. I Katsnelson, A. K Geim, I. V Grigorieva,
Nat. Comm. {\bf 4} 2010 (2013).
\bibitem{pike} N. A. Pike and D. Stroud
Phys. Rev. B {\bf 89} 115428 (2014).
\bibitem{Yazyev2007} O.\ V.\ Yazyev  and L.\ Helm,
 Phys.\ Rev.\  B {\bf 75}, 125408 (2007).
\bibitem{Rakhmanov12}A.\ L.\ Rakhmanov,  \ A.\ V.\  Rozhkov, A.\ O.\  Sboychakov, and F.\ Nori. 
Phys.\ Rev.\ {\bf B} {\bf 85}, 035408 (2012).
\bibitem{nakada} K.\ Nakada and A.\ Ishii, 
Solid State Commun.\ {\bf 151}, 13 (2011).
\bibitem{chan} K.\ T.\ Chan, J.\ B.\  Neaton, and M.\ L.\ Cohen, 
Phys.\  Rev.\  B {\bf 77}, 235430 (2008).
\bibitem{anderson1961} P.\ W.\ Anderson,
 Phys.\ Rev.\ {\bf 124}, 41 (1961).
\bibitem{pariser} R.\ Pariser and R.\ Parr, 
J.\ Chem.\ Phys.\ {\bf 21}, 767-775 (1953).
\bibitem{lykk} K.\ R.\ Lykke, K.\ K.\ Murray, and W.\ C.\  Lineberger, 
 Phys.\ Rev.\ {\bf A} {\bf 43}, 6104-6107 (1991).
\bibitem{wehling} T.\ O.\ Wehling, M.\ I.\ Katsnelson, and A.\ I.\ Lichtenstein, 
 Chem.\ Phys.\ Lett.\ {\bf 476} 125 (2009).
\bibitem{Hobson1953} J.\ Hobson and W.\ A.\  Nierenberg,
Phys.\ Rev.\ {\bf 86}, 662 (1953).
\bibitem{Yuan2010} S.\ Yuan, H.\  De Raedt, and M.\ I.\  Katsnelson, 
 \ Phys.\ Rev.\ {\bf B} {\bf 82}, 115448 (2010).
\end{thebibliography}
\end{document}